\documentclass[
aps,prd,
12pt,
nopreprintnumbers,
showpacs,
eqsecnum,
nofootinbib
]{revtex4-1}

\usepackage{graphicx}
\usepackage{amssymb}

\begin{document}

\title{
Quantum f{}luctuation of stress tensor in a higher-derivative scalar
f{}ield theory around a cosmic string}
\author{Nahomi Kan}\email[]{kan@gifu-nct.ac.jp}
\affiliation{National Institute of Technology, Gifu College,
Motosu-shi, Gifu 501-0495, Japan
}
\author{Masashi Kuniyasu}\email[]{mkuni13@yamaguchi-u.ac.jp}
\author{Kiyoshi Shiraishi}\email[]{shiraish@yamaguchi-u.ac.jp}
\affiliation{
Graduate School of Sciences and Technology for Innovation, Yamaguchi
University, Yamaguchi-shi, Yamaguchi 753--8512, Japan}
\date{\today}

\begin{abstract}
In this paper, we calculate the vacuum fluctuation of the stress tensor of a
higher-derivative theory around a thin cosmic string. 
To this end, we adopt the method to obtain the stress tensor from the
effective action developed by Gibbons \textit{et al.}
By their method, the quantum stress tensor of higher-derivative scalar theories
without self-interaction is expressed as a simple sum of quantum stress tensors of
free massive scalar fields.
Unlike the vacuum expectation value of the
scalar field squared obtained in the similar model, there appears no reduction of
the values near the conical singularity.
\end{abstract}



\maketitle

\section{Introduction}

The cosmic string \cite{Vilenkin,VS} is one of topological defects expected to
appear in phase transitions in the very early universe. It has been actively
researched as one related to the evolution and structure formation in the early
universe. A theoretically interesting thing is that, in the limit of thin
strings, the space-time around the cosmic string is locally flat
\cite{Vilenkin,VS} except for a conical singularity. The behavior of quantum fields
in the space-time with such a conical singularity, as a space with nontrivial
topology realized in our universe, has been extensively studied until now
\cite{HK,Linet,FS,Dowker1,Dowker2,Smith,AE,SH,CKV,GL,Moreira,FP}.
 
The study of
quantum fields in the vicinity of singularity may
reveal the 
essential
feature of quantum gravity, which is
incompletely defined at the present time. In particular, the consequence of
theoretical models in nontrivial space-time may give some hints for
quantum physics in strong gravity.
 
Now, there is a nontrivial expectation that the problems of divergence in 
quantum field theory in the ultraviolet region will be solved by incorporating
quantum gravity. 
Some kind of quantum-gravitational correction is
considered to be added to the standard quantum field theory,
and
a number of models
with nonlocality or higher-derivative operators has been studied
\cite{Modesto,BT,AP,Anselmi,BLM,BLMTY}.
 
We consider the model that Frolov and others first
proposed \cite{FS1,FS2,Fujikawa}, including an infinite number of derivatives in
the present paper. They originally considered such a model for the regularization
of physical quantities in field theory. Therefore, the divergence behavior related
to interacting fields is expected to be improved. 
 In a previous paper \cite{KKSW1}, we calculated the vacuum expectation
value of a scalar field squared of the higher-derivative theories
in a space with a conical singularity, which is
equivalent to that around an ideal cosmic string. It was confirmed that its
behavior at a short distance from the singularity was milder than that of the
canonical scalar model.
However, it is generally not
clear whether divergence cancels out for the physical quantity that is directly
connected to gravity, such as the quantum fluctuation of the stress tensor.
 
In this paper, we calculate the vacuum expectation value (in other words, quantum
fluctuation) of the stress tensor in the same background space-time. 
We adopt the method to obtain the stress tensor from the effective action
developed by Gibbons \textit{et al.}\cite{GPS},
and we investigate the characteristics of this physical
quantity in the present model.

The organization of the present paper is as follows.
In Sec.~\ref{sec2}, we formulate the method to obtain vacuum expectation values of
the stress tensor in the effective theory.
In Sec.~\ref{sec3}, we numerically evaluate the vacuum expectation values of
the stress tensor in the higher-derivative scalar field theory in a conical
space.
The results are summarized in Sec.~\ref{conclusion}.

\section{The expression for the quantum fluctuation of the stress tensor in the
higher-derivative theory}
\label{sec2}

In the present paper, we consider the following Lagrangian with higher derivatives
on a real scalar field $\phi$:
\begin{equation}
\mathcal{L}=-\frac{1}{2}\phi(x)\frac{\sqrt{-\Box}\sinh[\pi l\sqrt{-\Box}]}{{\pi}l}
\phi(x)\,,
\label{ThisL}
\end{equation}
where $\Box\equiv\nabla_\lambda\nabla^\lambda$ denotes the d'Alembertian
and $\nabla_\lambda$ is the covariant derivative.
Note that this reduces to the canonical massless scalar field Lagrangian if the
``cutoff length'' $l$ becomes zero. The Green's function in momentum space of this
model first appeared in Refs.~\cite{FS1,FS2,Fujikawa}. The Green's function in
flat configuration space can be written as \cite{KKSW1}%
\footnote{Incidentally, we need to consider an Euclidean space for obtaining the
vacuum expectation values from the Green's function.}
\begin{equation}
G(x,x')=\frac{l}{\pi^{\frac{D-1}{2}}}\sum_{k=0}^\infty
\frac{\Gamma\left(\frac{D-1}{2}\right)}{\left[|x-x'|^2+4\pi^2l^2
\left(k+\frac{1}{2}
\right)^2\right]^{\frac{D-1}{2}}}\,,
\end{equation}
where $\Gamma(s)$ denotes the gamma function.
It can be confirmed that the limit $l\rightarrow 0$ reduces this expression to the
standard massless Green's function.

The Lagrangian (\ref{ThisL}) can be written by using the infinite product as
\begin{equation}
\mathcal{L}=-\frac{1}{2C}\phi(x)(-\Box)\prod_{k=1}^{\infty}\left[
-\Box+\frac{k^2}{l^2}\right]\phi(x)\,,
\end{equation}
where $C=\prod_{k=1}^\infty\frac{k^2}{l^2}$.
Further, in order to apply the treatment according to Ref.~\cite{GPS},
we rewrite this in the form
\begin{equation}
\mathcal{L}=-\frac{1}{2C}\phi(x)\left[\prod_{i=1}^{\infty}A_i\right]\phi(x)\,,
\end{equation}
where $A_i=-\Box+m_i^2$, $m_i^2=(i-1)^2/l^2~(i=1, 2, 3, \ldots)$. Note that $m_1=0$
in the present model.

Gibbons \textit{et al.}\cite{GPS} recently constructed the stress
tensor of general higher-derivative theories from their effective Lagrangian.
The effective Lagrangian with multiple fields reads
\begin{equation}
\mathcal{L}=-\frac{1}{2C}\sum_{k=0}^\infty\eta_k(x)A_{k+1}\chi_{k+1}(x)
+\frac{1}{2C}\sum_{k=1}^\infty\eta_k(x)\chi_k(x)
\,,
\end{equation}
where $\eta_0\equiv\phi$ and $\chi_\infty\equiv\phi$. We further obtain the
relations, $\eta_k=\left[\prod_{i=1}^kA_i\right]\phi~(k\ge 1)$, and $\chi_k=
\left[\prod_{i=k+1}^\infty A_i\right]\phi~(k\ge 1)$ from the iterative use of
the equations of motion. In the present case, the number of fields is infinite.

Then, the stress tensor of the theory is given by \cite{GPS}
\begin{equation}
T_{\mu\nu}=\frac{1}{C}\sum_{k=0}^\infty\left[
\partial_{(\mu}\eta_k\partial_{\nu)}\chi_{k+1}-\frac{1}{2}g_{\mu\nu}
(g^{\rho\lambda}\partial_{\rho}\eta_k\partial_{\lambda}\chi_{k+1}+m_{k+1}^2\eta_k\chi_{k+1})\right]
+\frac{1}{2C}g_{\mu\nu}\sum_{k=1}^\infty\eta_k\chi_k\,,
\end{equation}
where $g_{\mu\nu}$ is the metric tensor of the background space-time.
From this expression, we can obtain the vacuum fluctuation of the stress tensor
$\langle T_{\mu\nu}\rangle$.
To estimate the quantum quantities, the fundamental basis is the Green's function%
\footnote{Because the Green's function is proportional to $\hbar$, the quantum
fluctuation we treated here is dubbed as the one-loop quantum effect.}
\begin{equation}
\langle\phi(x)\phi(x')\rangle=G(x,x')=\frac{C}{\prod_{i=1}^\infty A_i}\mathbf{1}
\,,
\end{equation}
where $\mathbf{1}$ denotes a covariant delta function
$\frac{1}{\sqrt{|g|}}\delta^D(x,x')$ in this symbolic expression.
From the Green's function and the relations to the original field $\phi$, it turns
out to be
\begin{equation}
\langle\eta_k(x)\chi_{k+1}(x')\rangle=
\left[\prod_{i=1}^kA_i\right]_x\left[\prod_{i=k+2}^\infty
A_i\right]_{x'}\langle\phi(x)\phi(x')\rangle=\frac{C}{
A_{k+1}}\mathbf{1}=C\Delta_{k+1}(x,x')
\,,
\end{equation}
where $\Delta_{k+1}(x,x')$ is the canonical Green's function of a scalar field
with mass $m_{k+1}$. Similarly, one finds
$\langle\eta_k(x)\chi_{k}(x')\rangle=C\mathbf{1}$.
Note also that
$m_{k+1}^2\langle\eta_k(x)\chi_{k+1}(x')\rangle-\langle\eta_k(x)\chi_{k}(x')\rangle
=\langle\eta_k(x)\Box\chi_{k+1}(x')\rangle$,
since
$\chi_k=A_{k+1}\chi_{k+1}$.

Consequently, we find a fairly simple expression for the quantum fluctuation of the
stress tensor:
\begin{equation}
\langle T^\rho_\sigma\rangle(x)=\lim_{x'\rightarrow x}\sum_{k=1}^\infty\left[
\nabla^{\rho}\nabla_{\sigma'}-\frac{1}{2}\delta_{\sigma}^{\rho}
(\nabla_{\lambda}\nabla^{\lambda'}+\nabla_{\lambda}\nabla^{\lambda})\right]
\Delta_k(x,x')
\,,
\end{equation}
where $\nabla_{\sigma}$ ($\nabla_{\sigma'}$) denotes the covariant derivative with
respect to the coordinates $x$ ($x'$).

In the following section, we evaluate the vacuum expectation value of the stress
tensor near a thin cosmic string using the expression obtained above.

\section{Numerical calculations of the quantum stress tensor around a conical
defect}
\label{sec3}

Now, we can express the vacuum
expectation value of the stress tensor, or the quantum stress tensor, in the
present model as
\begin{equation}
\langle T^\rho_\sigma\rangle(x)=\sum_{k=1}^{\infty} \langle
T^\rho_\sigma\rangle_k(x)\,,
\end{equation}
where
\begin{equation}
\langle
T^\rho_\sigma\rangle_k\equiv\lim_{x'\rightarrow
x}\left[\nabla^\rho\nabla_{\sigma'}
-\frac{1}{2}\delta^\rho_\sigma(\nabla^\lambda\nabla_{\lambda'}+
\nabla^\lambda\nabla_{\lambda})\right]
\Delta_k(x,x')\,,
\end{equation}
where
$\Delta_k(x,x')$ is the canonical Green's function of a free scalar field with mass
$m_k$. Note that the value of $\langle T^\rho_\sigma\rangle$ is still
a ``bare'' quantity (see, for example, Ref.~\cite{Smith}). 
When we consider a nontrivial background space-time, we should understand that
the physical quantities are the differences of the quantities from those obtained
in the trivial, flat and infinite space-times.

We examine the vacuum expectation values of the stress tensor in a conical space.
A conical space, or a space with a conical singularity at the coordinate origin,
is described by the metric
\begin{equation}
ds^2=\sum_{i=1}^{D-2}(dz^i)^2+dr^2+\frac{r^2}{\nu^2}d\theta^2\,,
\end{equation}
where $\nu$ is a constant greater than unity. This metric is equivalent to
\begin{equation}
ds^2=\sum_{i=1}^{D-2}(dz^i)^2+dr^2+r^2 d\tilde{\theta}^2\,,
\label{flat}
\end{equation}
where the range of $\tilde{\theta}$ is $0<\tilde{\theta}\le 2\pi/\nu$ 
($\nu\ge 1$).
This metric adequately describes a locally flat Euclidean space except for
the coordinate origin if $\nu\ne 1$. 
A Lorentzian space-time with a deficit angle is often
employed as a model space around a mathematically idealized straight cosmic
string \cite{Vilenkin,VS}. To obtain the quantum quantities in the space-time, we
have only to employ the space-time with the Euclidean signature.

The explicit form of the standard Green's function $\Delta_k(x,x')$ in a conical
space has already been known \cite{CKV,Moreira,SH,KKSW1,KKSW0}. 
To emphasize the difference from the flat space without the singularity, it is
natural to define
$\bar{\Delta}_k(x,x')\equiv\Delta_k(x,x')-\Delta_k(x,x')|_{\nu=1}$. Then, we find
\cite{KKSW1,KKSW0}
\begin{eqnarray}
& &\bar{\Delta}_k(x,x')=\bar{\Delta}_k(r,r',\varphi,\zeta)\nonumber \\
&=&\int_0^\infty ds\, {\textstyle 
\frac{e^{-\frac{r^2+{r'}^2+\zeta^2}{4s}-m_k^2s
}}{2\pi(4\pi
s)^{D/2}}}\int_{0}^\infty
e^{-\frac{rr'}{2s}\cosh v}\left[{\textstyle
\frac{\nu\sin\nu(\tilde{\varphi}-\pi)}{\cosh\nu
v-\cos\nu(\tilde{\varphi}-\pi)}-\frac{\nu\sin\nu(\tilde{\varphi}+\pi)}{\cosh\nu
v-\cos\nu(\tilde{\varphi}+\pi)}}\right]dv\,,
\label{sk}
\end{eqnarray}
where $x=(r,\theta, z^i)$, $x'=(r',\theta', z'^i)$, 
$\tilde{\varphi}=\tilde{\theta}-\tilde{\theta}'=\varphi/\nu=(\theta-\theta')/\nu$,
and $\zeta=\sqrt{(z^i-z'^i)^2}$. Using this ``renormalized'' Green's function, we
define the renormalized vacuum expectation value of the stress tensor $\langle
\bar{T}^\rho_\sigma\rangle$ as%
\footnote{At the present stage, one can explicitly check $\lim_{x'\rightarrow
x}(-\Box_x+m_k^2)\bar{\Delta}_k(x,x')=0$, and thus in this expression of the
renormalized stress tensor the former assumption of $m_1=0$ is irrelevant.}
\begin{equation}
\langle
\bar{T}^\rho_\sigma\rangle\equiv\sum_{k=1}^{\infty}\lim_{x'\rightarrow
x}\left[\nabla^\rho\nabla_{\sigma'}
-\frac{1}{2}\delta^\rho_\sigma(\nabla^\lambda\nabla_{\lambda'}+\nabla^\lambda
\nabla_{\lambda})\right]
\bar{\Delta}_k(x,x')\,.
\label{ef1}
\end{equation}

In this model, we thus find
\begin{eqnarray}
& &\sum_{k=1}^{\infty}\bar{\Delta}_k(x,x')\nonumber \\
&=&\int_0^\infty ds\,\varrho(s) {\textstyle 
\frac{e^{-\frac{r^2+{r'}^2+\zeta^2}{4s}
}}{2\pi(4\pi
s)^{D/2}}}\int_{0}^\infty
e^{-\frac{rr'}{2s}\cosh v}\left[{\textstyle
\frac{\nu\sin\nu(\tilde{\varphi}-\pi)}{\cosh\nu
v-\cos\nu(\tilde{\varphi}-\pi)}-\frac{\nu\sin\nu(\tilde{\varphi}+\pi)}{\cosh\nu
v-\cos\nu(\tilde{\varphi}+\pi)}}\right]dv\,,
\label{ef2}
\end{eqnarray}
where
\begin{equation}
\varrho(s)\equiv\sum_{k=0}^\infty
e^{-k^2s/l^2}=\frac{1}{2}\left[1+\vartheta_3(0,e^{-s/l^2})\right]
=\frac{1}{2}\left[1+l\sqrt{\frac{\pi}{s}}\vartheta_3(0,e^{-\pi^2l^2/s})\right]
\,,
\end{equation}
where $\vartheta_j(v,q)$ is the Jacobi theta function.

Now, we can evaluate the expectation values of the stress tensor in the present
higher-derivative theory, very efficiently using these expressions (\ref{ef1}) and
(\ref{ef2}).

\begin{figure}[ht]
\centering
\includegraphics[width=5cm]{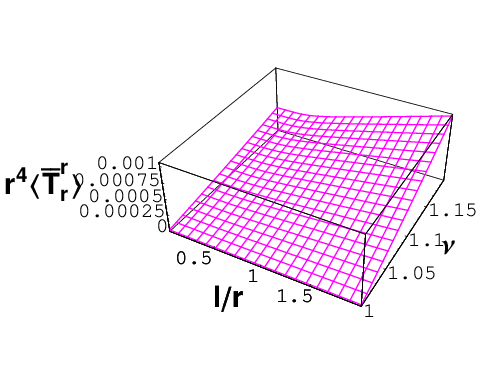}
\includegraphics[width=5cm]{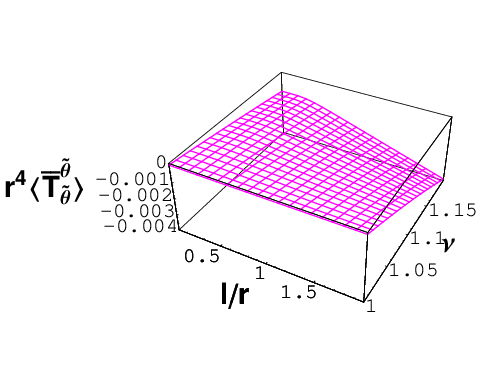}
\includegraphics[width=5cm]{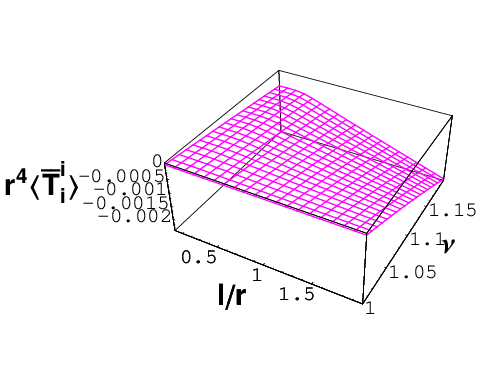}\\
\hspace{1cm} (a) \hspace{4.5cm} (b) \hspace{4.5cm} (c)
 \caption{The quantum stress tensor regulated as
$r^4\langle\bar{T}^\rho_\sigma\rangle$ for
$D=4$ are plotted against the cutoff length $l$ divided by the distance from the
origin $r$ in the present model, where (a) for $(\rho,\sigma)=(r,r)$,
(b) for $(\rho,\sigma)=(\tilde{\theta},\tilde{\theta})$, and (c) for
$(\rho,\sigma)=(i,i)$.}
\label{fig1}
\end{figure}

In Fig.~\ref{fig1}, we illustrate the $l/r$ dependence of the quantum
stress tensor in the present model for $D=4$. The limit of $l=0$ gives
the standard result with a canonical massless scalar field
\cite{HK,Linet,FS,Dowker1,Dowker2,Smith,AE,SH,CKV,GL,Moreira}:
$r^4\langle\bar{T}^r_r\rangle=
-\frac{1}{3}r^4\langle\bar{T}^{\tilde{\theta}}_{\tilde{\theta}}\rangle
=\frac{(\nu^2-1)(\nu^2+11)}{1440\pi^2}$ and $r^4\langle\bar{T}^i_i\rangle=
-\frac{(\nu^2-1)(19-\nu^2)}{1440\pi^2}$.
The correction due to the scale $l$ has significant dependence for a fixed
$r$. The absolute values of the components of the quantum stress tensor become
larger for larger $l$. This behavior is due to the summation form of the stress
tensor, in which there is no cancellation as observed in the vacuum expectation
value of the scalar squared \cite{KKSW1}.
Since the contributions from the massive modes exponentially decay at large $r$,
asymptotic expressions for the expectation value of the stress tensor are
$\langle\bar{T}^r_r\rangle=
-\frac{1}{3}\langle\bar{T}^{\tilde{\theta}}_{\tilde{\theta}}\rangle
\rightarrow\frac{(\nu^2-1)(\nu^2+11)}{1440\pi^2}/r^4$ and
$\langle\bar{T}^i_i\rangle\rightarrow -\frac{(\nu^2-1)(19-\nu^2)}{1440\pi^2}/r^4$
at large $r$.

\begin{figure}[ht]
\centering
\includegraphics[width=5cm]{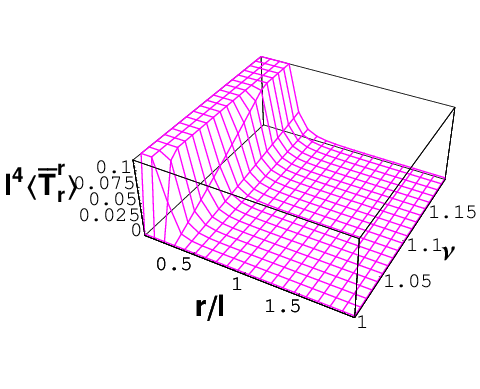}
\includegraphics[width=5cm]{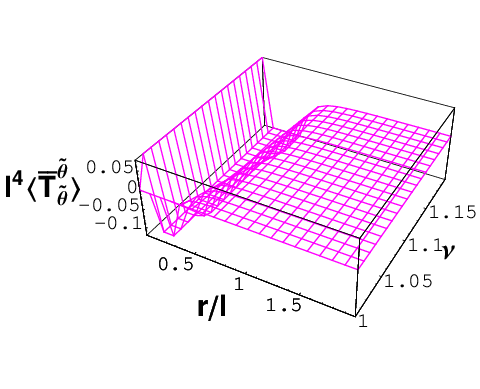}
\includegraphics[width=5cm]{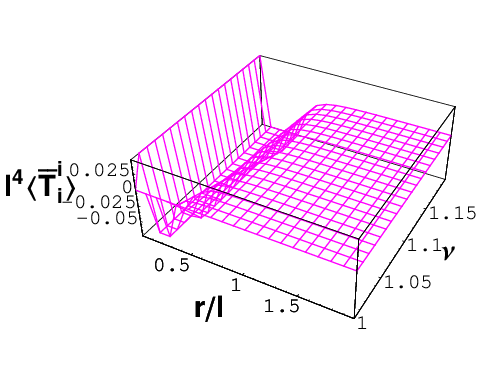}\\
\hspace{1cm} (a) \hspace{4.5cm} (b) \hspace{4.5cm} (c)
 \caption{The quantum stress tensor regulated as
$l^4\langle\bar{T}^\rho_\sigma\rangle$ for
$D=4$ are plotted against $r/l$ in the present model, where (a) for
$(\rho,\sigma)=(r,r)$, (b) for $(\rho,\sigma)=(\tilde{\theta},\tilde{\theta})$,
and (c) for
$(\rho,\sigma)=(i,i)$. Note that these values diverge near $r/l\sim 0$,
the numerical values near $r/l$ are expressed up to a finite cut in the figures.}
\label{fig2}
\end{figure}

In Fig.~\ref{fig2}, we show the $r/l$ dependence of the quantum
stress tensor in the present model.
Note that, in Fig.~\ref{fig2}, the values near $r=0$ is cut by a certain
finite value. The values at $r=0$ are still infinite as in the case with the
canonical scalar
field \cite{HK,Linet,FS,Dowker1,Dowker2,Smith,AE,SH,CKV,GL,Moreira}.

\section{Conclusion}
\label{conclusion}

In this paper, we calculated the quantum fluctuations of the stress tensor of a
higher-derivative scalar field model \cite{FS1,FS2,Fujikawa} in a conical space
which looks like a space-time around a cosmic string \cite{Vilenkin,VS}. We adopted
the stress tensor based on the effective theory of Gibbons \textit{et
al.}\cite{GPS}. In general, the quantum stress tensor of higher-derivative scalar
theories without self-interaction is expressed as a simple sum of quantum stress
tensors of free massive scalar fields. Therefore, it was found that there is no
cancellation that would appear in the behavior of Green's function and the
vacuum expectation value of a scalar field squared \cite{KKSW1}; some divergences
in canonical scalar field theory do not appear in the theory with infinite
derivatives. Numerical calculations reveal the dependence of the quantum
fluctuation of the stress tensor on the parameter $l$ corresponding to the cutoff
length included in the theory. It was also confirmed that results coincide with
those obtained with the canonical massless scalar field theory
\cite{Smith} in the limit of zero cutoff.
 
When we introduce self-interaction or interactions of multiple fields,
the quantum aspect of higher-derivative theories becomes completely nontrivial.
This is because the form of effective interaction becomes inevitably complicated
due to the redefinition of the field in effective action. However, if we
proceed with utilizing symmetries such as gauge symmetry, we may be able to study
physical quantities in certain restricted theories. In addition, it is worthwhile
to consider the compact space defined by the boundary condition on fields, where
we can perform calculations with considerably ease.

\bibliographystyle{apsrev4-1}

\end{document}